\documentclass[preprint,authoryear,12pt]{elsarticle}

\usepackage{graphics}

\usepackage{amssymb}

\journal{Journal of Atmospheric and Solar-Terrestrial Physics}

\begin{document}

\begin{frontmatter}



\title{Analyzing propagation of low-frequency dissipative oscillations in the upper atmosphere}

\author[label1]{G.V. Rudenko}
\ead{rus@iszf.irk.ru}

\address{Institute of Solar-Terrestrial Physics SB RAS, Irkutsk, Russia}

\begin{abstract}
At a horizontally homogeneous isothermal atmosphere approximation,
we derive an ordinary six-order differential equation describing
linear disturbances with consideration for heat conductivity and
viscosity of medium. The wave problem may be solved analytically
by representing the solution through generalized hypergeometric
functions only at a nonviscous heat-conducting isothermal
atmosphere approximation. The analytical solution may be used to
qualitatively analyze propagation of acoustic and internal gravity
waves  (AGWs) in the real atmosphere: a) to classify waves of
different frequencies and horizontal scales according to a degree
of attenuation  and thus according to their ability to appear in
observations and in general dynamics of the upper atmosphere; b)
to describe variations in amplitude and phase characteristics of
disturbances propagating in a height region with dominant
dissipation; c) to analyze  applicability of quasi-classical wave
description to a medium with exponentially growing dissipation. In
this paper, we also present wave and quasi-classical methods for
deriving waveguide solutions (dissipative ones corresponding to a
range of internal gravity waves (IGWs)) with consideration of wave
leakage into the upper atmosphere. We propose a qualitative scheme
which formally connects the wave leakage solution to the wave
solution in the upper dissipative atmosphere.  Spatial and
frequency characteristics of dissipative disturbances generated by
a waveguide leakage effect in the upper atmosphere are
demonstrated to agree well with observed characteristics of
middle-scale traveling ionospheric disturbances (TIDs).

\end{abstract}


\end{frontmatter}


\section{Introduction}\label{section1}
Acoustic gravity waves (AGWs) in the upper atmosphere cannot be
described without an accurate account of dissipative
characteristics of the environment. The dissipation effect on
structure and propagation of disturbances in the upper atmosphere
plays a very special role due to an exponential growth of its
kinematic value with height caused by an exponential decrease in
density. There is nothing to restrain the growth in wave
equations, which causes the dissipation to become, sooner or
later, dominant in a wave process, changing its physical
properties radically. Such a dissipation effect severely restricts
the use of geometrical optics (quasiclassical, WKB) approximation,
the most effective means of describing disturbances in stratified
media. This approximation is conditionally valid only to a
"critical" height $z_c$ at which ordinary nondissipative terms of
a wave system of equations with dissipative terms are
quantitatively compared. At heights of order of $z_c$ and higher,
the quasiclassical description is impossible at all. As a result,
a large number of disturbances in the upper atmosphere and their
associated ionospheric effects cannot be described and understood
in WKB. Thus, in response to dissipation, disturbances with small
wave vertical scales should be trivially reduced to vanishingly
small values at a height of order of $z_c$, and should not
significantly contribute to the atmospheric dynamics above this
height. In contrast, typical vertical scales of disturbances in
the upper atmosphere are generally comparable with the height of
the atmosphere. Their occurrence at large heights may be
understood only in the context of a rigorous wave description.
Being able to penetrate deeply into the upper atmosphere, such
disturbances may serve as agents transforming wave energy from
lower atmospheric layers into heat energy and thus may contribute
greatly not only to the dynamics but also to the general heat
balance of the upper atmosphere.

This paper relies on a description of wave disturbances in the
dissipative atmosphere in the form of solutions of a geometrical
optics equation. Fundamental aspects of this theoretical
description have been put forward before in \cite{Lyons};
\cite{Yanowitch_a}; \cite{Yanowitch_b} and have been stated in the
most common form in \cite{Rudenko_a}; \cite{Rudenko_a}. The
approach in hand ignores many aspects of the real atmosphere:
horizontal irregularity, wind motions, nonisothermality;
simultaneous overall consideration of dissipation (thermal
conductivity and viscosity) is impossible. The quasiclassical
approach does not have such limitations, but it is valid, as
indicated above, only to certain heights. Despite its
idealization, the rigorous wave description allows us to
understand the nature of dissipative disturbances in the upper
atmosphere, at least qualitatively.

The structure of this paper is as follows. Section \ref{section2}
gives a detailed derivation of a general wave equation in the form
of a six-order ordinary differential equation with one variable
comprising all types of dissipation: thermal conductivity and two
components of viscosity. The equation describes a vertical
structure of linear oscillations in the isothermal dissipative
atmosphere for arbitrary values of frequency $\omega$ and
horizontal wave number $k_x$. At low heights, where dissipation
can be ignored, this equation describes either acoustic or
internal-gravity oscillations, depending on the wave parameters.
In particular, this general equation readily yields all particular
types of wave description reducing to hypergeometric equations
discussed in  \cite{Lyons}; \cite{Yanowitch_a};
\cite{Yanowitch_b}; \cite{Rudenko_a}. The general equation is not
hypergeometric but, theoretically, can be solved numerically or
used for analyzing the asymptotic behavior of its solutions (see
\cite{Rudenko_b}). Section \ref{section3} discusses a fundamental
hypergeometric equation derived in \cite{Rudenko_a}. It is a
particular case of the general equation in which the first and
second viscosities are set equal to zero. In this case, the order
of the general equation decreases by 2, and its solutions are
represented by four independent generalized hypergeometric Meijer
$G$-functions. In this section, asymptotic properties of
independent solutions are used to find a unique meaningful
solution satisfying a finite condition in the upper half-space.
Formally, the solution describes full continua of acoustic and
gravity oscillations in a thermal conductivity medium. In the real
atmosphere, the Prandtl number is not sufficiently small to
justify the exclusion of dissipative terms of viscosity from
consideration. Nevertheless, the generality of the solution and
the rigor of the description totally justify the use of this
approximation for analyzing dissipative propagation of real
disturbances. Section \ref{section4} analyzes general properties
of penetration of waves with various periods and spatial scales
into the upper atmosphere and the degree of their attenuation
there. We present the $z_c$ dependence on an oscillation period in
the real atmosphere. This height conditionally divides the space
into two parts. In the lower half-space, the WKB wave description
is possible. In the upper half-space, dissipation is a main agent
determining a wave process. Seeing that dissipation strongly
suppresses waves with low vertical scales (relative to the scale
of the height of the atmosphere), $z_c$ may also be considered a
threshold height above which the said waves may be ignored in
physical considerations. To estimate the ability of arbitrary
acoustic and internal gravity waves to penetrate into the upper
atmosphere (above $z_c$), we propose to use the dependence of
characteristic of the degree of attenuation on wave period and
length. We show that the characteristics of the degree of
attenuation obtained from analytical and WKB solutions at the
limit of their co-validity correspond to each other. Section
\ref{section5} explores the possibility of using the analytical
solution, derived in Section \ref{section3}, for interpreting
upper-atmosphere disturbances of special range which can manifest
themselves in observations as traveling ionospheric disturbances
(TIDs) owing to ion-neutral collisions. Upward propagating
disturbances are considered to result from penetration of a
horizontally propagating normal IGW mode, located in the
stratosphere, from the waveguide. Using the NRLMSISE-2000
particular model atmosphere, we demonstrate that there may be only
one $0$-mode located at stratospheric heights. The mode dispersive
dependence of its period on a horizontal wavelength can be found
in two ways: through the WKB approximation and through the
numerical solution of a boundary wave problem. We estimate the
total change of wave amplitude during wave propagation through a
barrier opaque region and a propagation region to $z_c$ with due
regard to dissipation. The structure of the amplitude and phase
wave characteristics above $z_c$ is represented by an analytical
dissipative solution in the isothermal atmosphere. Section
\ref{section6} summarizes findings of this study and discusses
possibilities of their application.
\section{Equation of dissipative linear oscillations in the isothermal atmosphere}\label{section2}
Consider the isothermal atmosphere with constant coefficients of
thermal conductivity and viscosity, which is formally determined
in the whole space:
\begin{equation}\label{eq1}
\begin{array}{l}
  p_0(z)=p_0(z_r)e^{-\frac{z-z_{r}}{H}}, \\
  \rho_0(z)=\rho_0(z_r)e^{-\frac{z-z_{r}}{H}}, \\
  H=\frac{RT_0}{g}, \kappa=const, \nu_1=const, \nu_2=const.
\end{array}
\end{equation}

Here $p_0, \rho_0$ are undisturbed density and pressure  of a
medium; $H$ is the height of the atmosphere; $R$ is the gas
constant; $g$ is the free fall acceleration;  $\kappa, \nu_1,
\nu_2 $ are dynamic coefficients of thermal conductivity, first
and second viscosities respectively; $z_r$ is the reference height
with specified undisturbed pressure and density. Without loss of
generality, we will assume that all disturbances are independent
of the $y$-coordinate and take the form: $Q'=Q'(z)e^{-i\omega t}$.
For the convenience of the wave description, we will exploit a
completely dimensionless form of its representation, i.e.
coordinates, time, wave parameters, and disturbance function will
be represented by corresponding dimensionless values:
 \begin{equation}\label{eq2}
\begin{array}{l}
  {\bf r}^*\equiv(x^*,y^*,z^*)={\bf r}/H\equiv(x,y,z)/H, t^*=t\sqrt{g/H},  \\
  k=k_xH, \sigma=\omega\sqrt{H/g}, \\
  n=\rho'/\rho_0, f=p'/p_0, \Theta=T'/T_0, \\
  u=v_x/\sqrt{Hg}, w=v_y/\sqrt{Hg}, v=v_z/\sqrt{Hg}.
\end{array}
\end{equation}
To bring the disturbance equations to dimensionless form, we will
use dimensionless expressions of kinematic dissipative values:
\begin{equation}\label{eq3}
\begin{array}{l}
  s(z)=\frac{\kappa}{\sigma\gamma c_vH\sqrt{gh}}\rho_0^{-1}, \\
  \mu(z)=\frac{\nu_1}{\sigma H\sqrt{gh}}\rho_0^{-1}, \\
  q(z)=\frac{\nu_1/3+\nu_2}{\sigma H\sqrt{gh}}\rho_0^{-1}.
\end{array}
\end{equation}
Here $\gamma$ is the adiabatic index; $c_v$ is the specific heat
capacity at constant volume.  By applying introduced
determinations in Eqs. (\ref{eq1})-(\ref{eq3}) to linearized
equations of state, continuity, moment, and entropy change, we
obtain a complete system of equations for disturbances of density,
pressure, velocity, and temperature:
\begin{equation}\label{eq4}
\begin{array}{l}
  a) \ \ \ \ \Theta=f-n, \\
  b) \ \ \ \ -i\sigma n+\Psi-v=0, \\
  c) \ \ \ \ -i\sigma f-v+\gamma\Psi=\sigma\gamma s\Delta\Theta, \\
  d) \ \ \ \ -i\sigma u+ikf=\sigma\mu\Delta u+ik\sigma q\Psi, \\
  e) \ \ \ \ -i\sigma v+\dot{f}-f+n=\sigma\mu\Delta v+\sigma q\dot{\Psi}, \\
  f) \ \ \ \ i\sigma w=\sigma\mu\Delta w.
\end{array}
\end{equation}
Here, the dot is the derivative of a function with respect to
$z^*$ argument; $\Psi=\dot{v}+iku$ is the dimensionless divergence
of velocity disturbance; $\Delta=d^2/{dz^*}^2-k^2$ is the
dimensionless Laplacian. Eq. (\ref{eq4}f) describes an independent
trivial viscous solution unrelated to the disturbances we are
interested in. Thus, from now on we set $w=0$ and will consider
the system of five Eqs. (\ref{eq4}a)-(\ref{eq4}e) with unknowns
($\Theta, n, f, u, v$). The kinematic, dissipative coefficients
$s$, $\mu$, and $q$ are functions which grow exponentially with
height. At low heights, where these coefficients may be ignored,
equation system of Eqs. (\ref{eq4}) describes ordinary classical
acoustic and gravitational oscillations. In the real atmosphere,
thermal conductivity exceeds viscosity. Hence, with an increase in
height, thermal-conductivity dissipation does occur first, i.e.
where the dimensionless function $s$ becomes of order of unity. A
corresponding height in the real atmosphere may be determined
using first formula of Eqs. (\ref{eq3}) from the condition
 \begin{equation}\label{eqd1}
\frac{\kappa}{\omega\gamma c_v\left(H(z_c)\right)^2\rho_0(z_c)}=1.
\end{equation}
Here, the height dependence of the height of the atmosphere is
taken into account. In what follows, we show that the presence of
the wave frequency $\omega$ in this condition is due to properties
of wave solutions  in which lower and upper asymptotic behaviors
depend on the ratio of $s$ to unity. The value of $z_c$ may be
found by solving implicit Eq. (\ref{eqd1}). Reference to Eq.
(\ref{eqd1})) shows that with an increase in frequency of
oscillations the critical height should also increase. Next, we
will assume for convenience that the point of reference of the
dimensionless coordinate $z$ corresponds to $z_c$ :
 \begin{equation}\label{eqd2}
z^*=(z-z_c)/H.
\end{equation}
Accordingly, $p_0(z_r)$ and $\rho_0(z_r)$ in Eq.(\ref{eq1}) are
determined by values at $z_c=z_r$ and, therefore, are implicit
functions of frequency of oscillations too.  From Eqs.
(\ref{eqd1}) and (\ref{eqd2}), expressions for $s$, $\mu$, and $q$
take the following form:
\begin{equation}\label{eqd3}
\begin{array}{l}
  s=e^{z^*}, \\
  \mu=\frac{\nu_1}{\kappa}c_ve^{z^*}, \\
  q=\frac{\nu_1/3+\nu_2}{\kappa}c_ve^{z^*}.
\end{array}
\end{equation}

Equation system (\ref{eq4}a)-(\ref{eq4}e) allows reducing to one
six-order ordinary differential equation with one variable
$\Theta$ . To derive this equation, we exclude variables in the
following sequence. First, we exclude variables $n$ and $f$ from
Eqs. (\ref{eq4}a),(\ref{eq4}b):
  \begin{equation}\label{eq5}
n=\frac{1}{i\sigma}(\Psi-v),  \ \ \
f=\Theta+n=\Theta+\frac{1}{i\sigma}(\Psi-v)
\end{equation}
Using (\ref{eq5}), we obtain the following expressions:
\begin{equation}\label{eq6}
\begin{array}{l}
  a) \ \ \ \ \Psi=\frac{\sigma}{\gamma-1}(\gamma s\Delta\Theta+i\Theta),
  b) \ \ \ \ -i\sigma u+\frac{k}{\sigma}v+\sigma\mu\Delta u=ik\Theta-ik\left(\sigma q-\frac{1}{i\sigma}\right)\Psi, \\
  c) \ \ \ \ -i\sigma v+\frac{1}{i\sigma}\dot v+\sigma\mu\Delta v=\dot\Theta-\Theta-\left(\sigma q-\frac{1}{i\sigma}\right)\dot\Psi, \\
  d) \ \ \ \ \Psi=\dot{v}+iku, \\
  e) \ \ \ \ \theta=\dot u-ikv, \\
  f) \ \ \ \ \Delta u=\dot\theta+ik\Psi, \\
  g) \ \ \ \ \Delta v=\dot\Psi+ik\theta.
\end{array}
\end{equation}
Here, Eq. (\ref{eq6}a) results from the subtraction of Eq.
(\ref{eq4}b) from Eq. (\ref{eq4}c) with the use of Eq.
(\ref{eq4}a); Eqs. (\ref{eq6}b) and (\ref{eq6}c) result from the
substitution of Eq. (\ref{eq5}) into Eq. (\ref{eq4}d) and Eq.
(\ref{eq4}e) respectively; Eq. (\ref{eq6}d) is the determination
of divergence; Eq. (\ref{eq6}e) is a new auxiliary function of
current; Eq. (\ref{eq6}f) and Eq. (\ref{eq6}g) are auxiliary
identities evident from the determinations of divergence and
current function. Eq. (\ref{eq6}à) gives us an explicit expression
of divergence through temperature disturbance $\Theta$. The
current function may also be explicitly expressed through $\Theta$
by summing up differentiated Eq. (\ref{eq6}c) and Eq. (\ref{eq6}b)
multiplied by $ik$ :
   \begin{equation}\label{eq7}
\theta=\frac{\sigma}{k}\frac{1}{1+i\sigma^2\mu}\left\{
\hat{L}\left[\sigma(\mu+q)\Psi-\Theta-\frac{1}{i\sigma}\Psi\right]+i\sigma\Psi\right\},
\end{equation}
where $\hat{L}=\Delta-\frac{d}{dz^*}$. By differentiating Eq.
(\ref{eq6}b) and subtracting Eq. (\ref{eq6}c) multiplied by $ik$,
we derive a differential equation expressed through one unknown
function $\Theta$ :
    \begin{equation}\label{eq8}
\left(1-i\hat{L}\mu\right)\theta+k(\mu+q)\Psi-\frac{k}{\sigma}\Theta=0,
\end{equation}
or in more detail
    \begin{equation}\label{eq9}
\left(1-i\hat{L}\mu\right)\frac{\hat{L}\left[\sigma(\mu+q)\Psi-\Theta-\frac{1}{i\sigma}\Psi\right]+i\sigma\Psi}{1+i\sigma^2\mu}+\frac{k^2}{\sigma}(\mu+q)\Psi-\frac{k^2}{\sigma^2}\Theta=0.
\end{equation}
Other values are represented by $\Theta$ the following set of
expressions:
        \begin{equation}\label{eq10}
\begin{array}{l}
  v=\frac{\sigma^2}{\sigma^4-k^2}\times \\
   \left[\left(1-\frac{d}{dz^*}-\frac{k^2}{\sigma^2}\right)(\Psi+i\sigma\Theta)+i\sigma^2(\mu+q)\left(\frac{d}{dz^*}+\frac{k^2}{\sigma^2}\right)\Psi-k\mu\left(\frac{d}{dz^*}+\sigma^2\right)\theta   \right], \\
  f=\Theta+\frac{i}{\sigma}(v-\Psi), \\
  u=\frac{k}{\sigma}f+i\mu\dot{\theta}-k(\mu+q)\Psi,\\
  n=f-\Theta
\end{array}
\end{equation}
Expressions Eq.(\ref{eq9}) and Eqs. (\ref{eq10}) completely
describe the wave disturbance in the selected model of medium. Eq.
(\ref{eq9}) does not have analytical solutions, but it may be used
for analyzing the asymptotic behavior of solutions at large and
small $z^*$, or for numerical solution (see \cite{Rudenko_a};
\cite{Rudenko_b}). Unlike the classical dissipationless solution,
Eq. (\ref{eq9}) allows the solution without an infinite increase
in amplitude of relative values of disturbances; i.e., in the
whole space, the solution may satisfy a linear approximation
(\cite{Rudenko_a}).
\section{Rigorous solution for AGV in the isothermal heat-conducting atmosphere}\label{section3}
The most interesting possibility of deriving an analytical form of
dissipative solutions which describes disturbances of acoustic and
gravitational ranges is provided by a model of viscosity-free
heat-conducting medium ($\mu=q=0$ ). In this case, Eqs.
(\ref{eq9}) and (\ref{eq10}) take the following form:
    \begin{equation}\label{eq11}
\left[-\hat{L}\left(\Theta+\frac{1}{i\sigma}\Psi\right)+i\sigma\Psi\right]-\frac{k^2}{\sigma^2}\Theta=0;
\end{equation}

\begin{equation}\label{eq12}
\begin{array}{l}
  \Psi=\frac{\sigma}{\gamma-1}\left(\gamma e^{z^*}\Delta\Theta+i\Theta\right), \\
  v=\frac{\sigma^2}{\sigma^4-k^2}\left(1-\frac{d}{dz^*}-\frac{k^2}{\sigma^2}\right)(\Psi+i\sigma\Theta), \\
  f=\Theta+\frac{i}{\sigma}(v-\Psi),\\
  u=\frac{k}{\sigma}f, \\
  n=f-\Theta.
\end{array}
\end{equation}

Introducing a new variable
    \begin{equation}\label{eq13}
\xi=\exp\left(-z^*+i\pi\frac{3}{2}\right)
\end{equation}
allows us to represent Eq. (\ref{eq11}) in the canonical form of
generalized hypergeometric equation
   \begin{equation}\label{eq14}
\left[\xi\prod^2_{j=1}(\delta-a_j+1)-\prod^4_{i=1}(\delta-b_i)\right]\Theta=0,
\end{equation}
where  $\delta=\xi d/d\xi$, $a_{1,2}=\frac{1}{2}\pm iq$,
 $b_{1,2}=\pm k$,$b_{3,4}=1/2\pm\alpha$
 \begin{equation}\label{eq15}
\begin{array}{l}
  q=\sqrt{-\frac{1}{4}+\frac{\gamma-1}{\gamma}\frac{k^2}{\sigma^2}+\frac{\sigma^2}{\gamma}-k^2}, \\
  \alpha=\sqrt{\frac{1}{4}+k^2-\sigma^2}.

\end{array}
\end{equation}
Eq. (\ref{eq14}) has two singular points:  $\xi=0$  (the regular
singular point,  $z^*=+\infty$) and  $\xi=\infty$   (the irregular
singular point, $z^*=-\infty$)). The fundamental system of
solutions of Eq. (\ref{eq14}) may be expressed by four linearly
independent generalized Meijer functions (\cite{Luke}):
 \begin{equation}\label{eq16}
\begin{array}{l}
  a) \ \ \ \ \ \Theta_1=G^{4,1}_{2,4}\left(\xi e^{-i\pi}\left|^{a_1,a_2}_{b_1,b_2,b_3,b_4}\right.\right),\\
  b) \ \ \ \ \ \Theta_2=G^{4,1}_{2,4}\left(\xi e^{-i\pi}\left|^{a_2,a_1}_{b_1,b_2,b_3,b_4}\right.\right),\\
  c) \ \ \ \ \ \Theta_3=G^{4,0}_{2,4}\left(\xi e^{-2i\pi}\left|^{a_1,a_2}_{b_1,b_2,b_3,b_4}\right.\right), \\
  d) \ \ \ \ \ \Theta_4=G^{4,0}_{2,4}\left(\xi\left|^{a_1,a_2}_{b_1,b_2,b_3,b_4}\right.\right).
  \end{array}
\end{equation}
Desired meaningful solution of Eq. (\ref{eq11}) may be found from
known properties of asymptotic behaviors of $\Theta_i$-functions
at two singular points in Eq. (\ref{eq14}).

At the irregular point $\xi=\infty$ ($z^*=-\infty$), we have:
 \begin{equation}\label{eq17}
\begin{array}{l}
  a,b) \ \ \  \Theta_{1,2}(\xi)\sim\Theta_{1,2}^{\infty}(\xi)=p_{1,2}\cdot e^{(1/2\mp iq)z^*}=\\
  \ \ \ \frac{\Gamma\left(\frac{1}{2}\mp iq+k\right)\Gamma\left(\frac{1}{2}\mp iq-k\right)\Gamma(1\mp iq+\alpha)\Gamma(1\mp iq-\alpha)}
  {\Gamma(1\mp 2iq)}e^{\mp\frac{\pi q}{2}-i\frac{\pi}{4}}\cdot e^{(1/2\mp iq)z^*},\\
  \\
  c,d) \ \ \  \Theta_{3,4}(\xi)\sim\Theta_{3,4}^{\infty}(\xi)=\pi^{1/2}e^{-i\frac{\pi}{8}(1\mp 2)}\cdot e^{z^*/4}e^{\mp\sqrt{2}(1-i)e^{-z^*/2}}.

  \end{array}
\end{equation}
From Eqs. (\ref{eq17}a), (\ref{eq17}b) follows that, at real $q$,
asymptotics $\Theta_1$ and $\Theta_2$ correspond to two
differently-directed classical dissipationless waves. Here, we
will assume that $\Theta_1$ corresponds to an upward propagating
wave; $\Theta_2$, to a downward propagating wave. Then, for
$\sigma$ and $k$ corresponding to acoustic waves, we set  $q<0$;
for those corresponding to internal gravity waves,  $q>0$.
Asymptotics $\Theta_3$ and $\Theta_4$ specified by Eqs.
(\ref{eq17}c), (\ref{eq17}d) correspond to dissipative
oscillations. These asymptotics are functions of an extremely
rapid decrease and a downward increase in  $z^*$. Decreasing
$\Theta_4$ leaves the physical scene very quickly at decreasing
$z^*$, and an asymptotic growth of $\Theta_4$  provokes the
necessity to completely exclude $\Theta_4$ from the desired
meaningful solution.

The behavior of other solutions $\Theta_1$, $\Theta_2$, and
$\Theta_3$ nearby the regular point $\xi=0$ ($z^*=+\infty$) is
represented by the following expressions:
    \begin{equation}\label{eq18}
\Theta_i\sim\Theta_i^0=\sum^4_{j=1}t_{ij}e^{-b_jz^*} \ \ \
(i=1,2,3).
\end{equation}
 Expressions for $t_{ij}$ are derived from known asymptotics of the Meijer G-function at the regular singular point (\cite{Luke}).
 Their explicit expressions are listed in \ref{A}.

When constructing the meaningful solution, we will assume that
there should not be upward growing asymptotic terms $\sim
e^{-b_2z^*}$ and $\sim e^{-b_4z^*}$. Besides, we set the incident
wave amplitude $= 1$. Accordingly, the desired solution is found
in the following form
    \begin{equation}\label{eq19}
\Theta(\xi)=p_1^{-1}\Theta_1(\xi)+\alpha_2\Theta_2(\xi)+\alpha_3\Theta_3(\xi),
\end{equation}
where coefficients $\alpha_2$ and $\alpha_3$ are selected from the
condition of elimination of growing asymptotics at the regular
point of  Eq. (\ref{eq14}):
 \begin{equation}\label{eq20}
\begin{array}{l}
 p_1^{-1}t_{12}+\alpha_2t_{22}+\alpha_3t_{32}=0 ,\\
  p_1^{-1}t_{13}+\alpha_2t_{23}+\alpha_3t_{33}=0 .
  \end{array}
\end{equation}
Solving system of Eqs (\ref{eq20}) yields:
 \begin{equation}\label{eq21}
\begin{array}{l}
  a) \ \ \ \ \  \alpha_2=-p_1^{-1}e^{-2\pi q}\frac{\sin[\pi(\alpha-iq)]\cos[\pi(k-iq)]}{\sin[\pi(\alpha+iq)]\cos[\pi(k+iq)]},\\
  b) \ \ \ \ \  \alpha_3=2\pi p_1^{-1}\frac{e^{-\pi q}}{e^{i2\pi\alpha}+e^{i2\pi k}}\left\{\frac{\sin[\pi(\alpha-iq)]}{\sin[\pi(\alpha+iq)]}- \frac{\cos[\pi(k-iq)]}{\cos[\pi(k+iq)]}\right\}.

  \end{array}
\end{equation}
Eqs. (\ref{eq19}) and (\ref{eq21}a), (\ref{eq21}b) give a strict
analytical expression of the desired meaningful solution. For
$|\xi|<1$ ($z^*>0$), solution of Eq. (\ref{eq19}) may be expressed
through generalized hypergeometric functions $_mF_n$ which, in
this region of acceptable arguments, are represented by simple
convergent power series suitable for the numerical calculation.
Such a representation of solution is obtained from the standard
representation of the Meijer G-function.
  \begin{equation}\label{eq22}
\begin{array}{l}
  G^{m,n}_{p,q}\left(y\left|^{a_p}_{b_q}\right.\right)=\sum^m_{h=1}\frac{\prod^m_{j=1}\Gamma(b_j-b_h)^*\prod^n_{j=1}\Gamma(1-a_j-b_h)}{\prod^q_{j=m+1}\Gamma(1-b_h-b_j)\prod^n_{j=n+1}\Gamma(a_j-b_h)}y^{b_h}\times\\
  \ \ \  _pF_{q-1}\left(\left.^{1+b_h-a_p}_{1+b_h-b_q^*}\right|(-1)^{p-m-n}y\right).
  \end{array}
\end{equation}
Here (*) implies that a term with an index equal to $h$ is
omitted. With Eq. (\ref{eq22}), after the rather lengthy
calculations, we can reduce solution of Eq. (\ref{eq19}) to the
following form:
  \begin{equation}\label{eq23}
\begin{array}{l}
\Theta(z^*>0)=\beta_0\beta_1 e^{-kz^*}\times _2F_3\left(\left.^{\frac{1}{2}+k-iq,\frac{1}{2}+k+iq}_{1-2k,\frac{1}{2}+k+\alpha,\frac{1}{2}+k-\alpha}\right|ie^{-z^*}\right)+ \\
 \ \ \ \beta_0\beta_2 e^{-(1/2+\alpha)z^*}\times _2F_3\left(\left.^{1+\alpha-iq,1+\alpha+iq}_{\frac{3}{2}+\alpha-k,\frac{3}{2}+\alpha+k,1+2\alpha}\right|-ie^{-z^*}\right),
  \end{array}
\end{equation}
where \\
 $\beta_0=\frac{\Gamma\left(\frac{1}{2}+iq+k\right)\Gamma(iq+\alpha)}{\Gamma(2iq)\Gamma\left(\frac{1}{2}+\alpha+k\right)}e^{-i\frac{\pi}{4}-\pi\frac{q}{2}}$,
$\beta_1=\frac{\Gamma\left(\frac{1}{2}+\alpha-k\right)\Gamma\left(\frac{1}{2}+iq+k\right)}{\Gamma(1+2k)\Gamma(1-iq+\alpha)}e^{-i\frac{\pi}{2}k}$,\\
$\beta_2=\frac{\Gamma\left(-\frac{1}{2}-\alpha+k\right)\Gamma(1+iq+\alpha)}{\Gamma\left(\frac{1}{2}+\alpha+k\right)\Gamma(1+2\alpha)\Gamma\left(\frac{1}{2}-iq+k\right)}e^{-i\frac{\pi}{2}\left(\frac{1}{2}+\alpha\right)}$.
\\
Given $z^*\rightarrow \infty$, the generalized hypergeometric
functions $F$ in Eq. (\ref{eq23}) tend to unity, and the solution
$\Theta$ takes a simple asymptotic form with two exponentially
decreasing terms.

The computed solution describes the incidence of internal gravity
or acoustic wave with an arbitrary inclination to the dissipative
region  $z^*>0$, its reflection from this region, and its
penetration to this region with the transformation of it into a
dissipative form. The complex coefficient of reflection can be
expressed by a simple analytical expression:
      \begin{equation}\label{eq24}
K=\alpha_2p_2.
\end{equation}
In \cite{Rudenko_a}, the reflection coefficient module is shown to
take a value of order of unity, if the typical vertical scale of
incident wave $q^{-1}\gtrsim 1$. Otherwise, the contribution of
the reflected wave exponentially decreases with decreasing
vertical scale. This behavior of reflection is similar to the
ordinary wave reflection from an irregularity of a medium. In our
case, the scale of the irregularity is the height of the
atmosphere $H$. It is worth introducing one more value
characterizing a value of wave attenuation in the region
($z^*<0$):
      \begin{equation}\label{eq25}
\eta=|\Theta(0)|.
\end{equation}
This value may be calculated using the form of solution of Eq.
(\ref{eq23}). In what follows, we will show that the behavior of
$\eta$, depending on a vertical wave scale, is similar to the
behavior of reflection coefficient. The value $\eta$ also
characterizes the capability of a significant part of wave
disturbance to penetrate to the upper region $z^*>0$. It is
reasonable that, if $\eta$ is negligible, further wave disturbance
propagation may be neglected too.
\section{Classification of AGV in the real atmosphere by dissipative properties}\label{section4}
The understanding of certain wave effects at heights of the upper
atmosphere requires knowing possible modes of wave propagation
which are associated with dissipation effect. Of particular
interest first is to determine the typical height from which
dissipation assumes a dominant control over the wave process, and
second is the type of wave propagation, depending on the period
and spatial scale of disturbance (whether it is an approximately
dissipativeless propagation or propagation with dominant
dissipation).
\subsection{Classification of waves by the parameter of dissipation critical height}\label{section4.1}
The dependence of $z_c$ on oscillation frequency enables a
convenient classification of waves only by their oscillation
periods. To make such a classification, we will exploit the
following model of medium: \\
-- The vertical distribution of temperature $T_0(z)$ according to
the NRLMSISE-2000 distribution with geographic coordinates of
Irkutsk for the local noon of winter opposition; \\
-- $p_0(z)=p_0(0)\exp\left[-\frac{g}{R}\int_0^z\frac{1}{T(z')}dz'\right], \ \ \ p_0(0)=1.01 \rm{Pa}$; \\
--  $\rho_0(z)=\rho_0(0)\exp\left[-\frac{g}{R}\int_0^z\frac{1}{T(z')}dz'\right], \ \ \ \rho_0(0)=287.0 \rm{g/m^3}$; \\
-- $g=9.807\rm{m/s^2}$, $R=287\rm{J/(kg\cdot K)}$,
$\kappa=0.026\rm{J/(K\cdot m\cdot s)}$,\\
$c_v=716.72\rm{J/(kg\cdot K)}$.

\begin{figure}
    \includegraphics[width=1\linewidth]{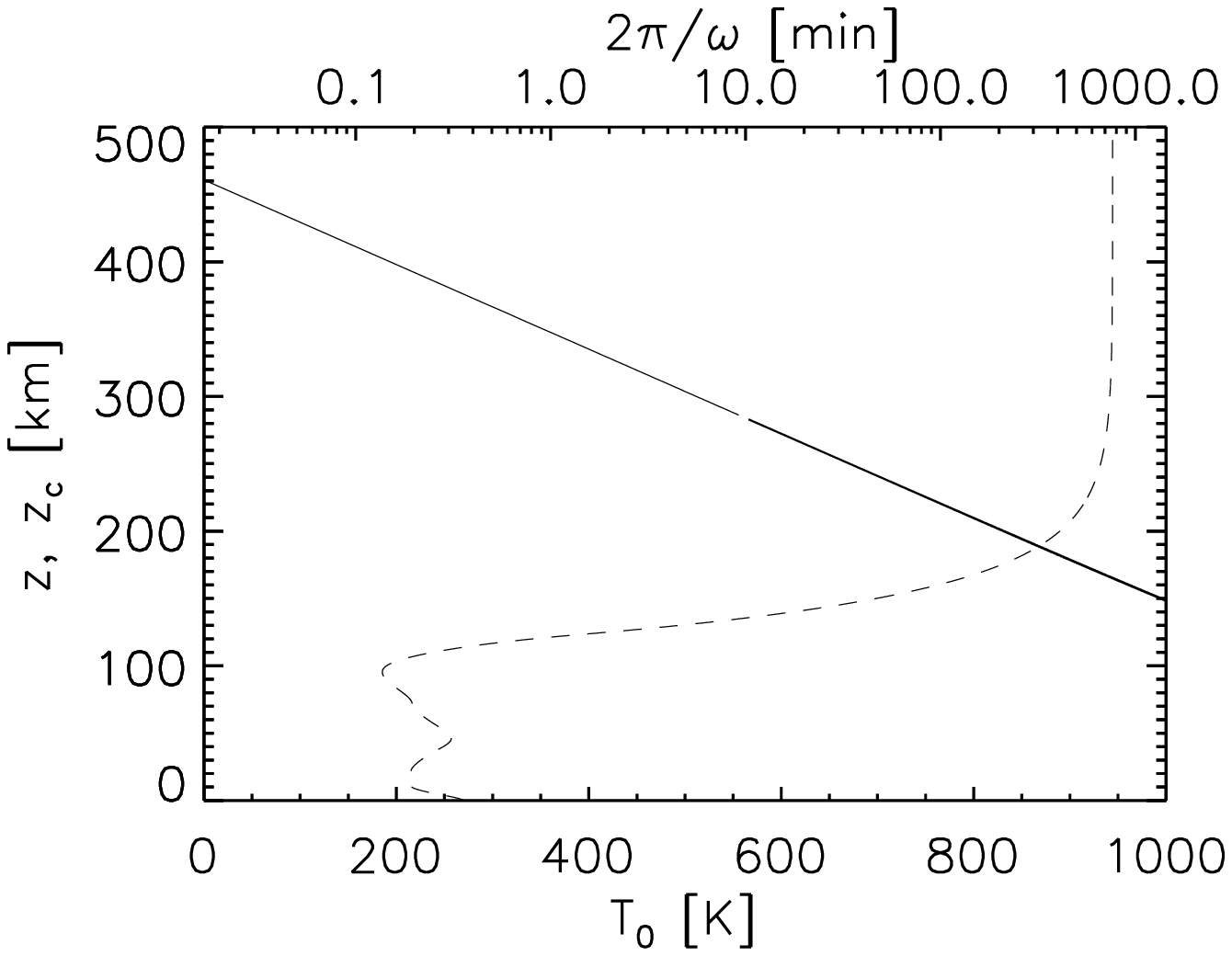}\\
  \caption{The dashed curve shows the height dependence of temperature in the selected model;
  the thick solid curve indicates the dependence of $z_c$ on the period of oscillations corresponding
  to IGW waves incident from the lower atmosphere; the solid curve indicates the dependence of $z_c$ on the period of oscillations corresponding
  to acoustic waves incident from the lower atmosphere.}\label{fig1}
\end{figure}

  Solving Eq. (\ref{eqd1}) for $z_c$ (the plot in Fig.  \ref{fig1}) provides the following useful information:
  a) the height of transformation of waves of the selected period into dissipative oscillations;
  b) the height above which waves of the selected period with vertical scales much less than the height of the atmosphere
  should be heavily suppressed by dissipation (in fact, such waves should not appear at these heights);
  c) the height limiting the applicability of the WKB approximation for the wave of the selected period.
\subsection{Classification of AGV by damping value}\label{section4.2}
An important characteristic of a wave with certain wave parameters
$\sigma$ and $k$ is the ratio of the amplitude of the wave
solution at $z_c$ to the amplitude of the wave incident from minus
infinity $\eta$, Eq. (\ref{eq25}). This ratio gives useful
information about the ability of this wave to have physically
meaningful values at heights of order of and higher than $z_c$. A
direct calculation of the wave solution for the height region
$z<z_c$ from general expression of Eq. (\ref{eq19}) cannot be
technically feasible. The first term of asymptotic expansion Eq.
(\ref{eq17}a) describes the solution only at a sufficiently large
distance from $z_c$ and  does not reflect the real behavior of the
amplitude under the influence of dissipation because this term has
the form of ordinary propagating dissipationless wave. Only
beginning from $z_c$, we can exactly calculate the solution by
power expansion Eq. (\ref{eq23}). On the other hand, the
characteristic we are interested in may be compared with its
values obtained in the WKB approximation for Eq. (\ref{eq14}). The
use of the WKB approximation also enables us to obtain the
qualitative behavior of the solution in the region $z\lesssim z_c
$. For simplicity, consider a case when a WKB estimated value
$\eta$ may be expressed in an analytical form. For this purpose,
rewrite Eq. (\ref{eq14}) for a new variable $\Theta_{ref}=\Theta
e^{-\frac{1}{2}z^*}$ :
      \begin{equation}\label{eq26}
\left\{\frac{d^2}{{dz^*}^2}+q^2+\frac{e^{z^*}}{i}\left(\frac{d^2}{{dz^*}^2}-\frac{1}{4}-k^2+\sigma^2
\right)\left[\left(\frac{d}{dz^*}+\frac{1}{2}\right)^2-k^2\right]\right\}\Theta_{ref}=0.
\end{equation}
In the WKB approximation, Eq. (\ref{eq26}) yields a quartic
algebraic equation for a complex value of a dimensionless vertical
wave number $k_z^*$:
      \begin{equation}\label{eq27}
-{k_z^*}^2+q^2+\frac{e^{z^*}}{i}\left(-{k_z^*}^2-\frac{1}{4}-k^2+\sigma^2
\right)\left[\left(ik_z^*+\frac{1}{2}\right)^2-k^2\right]=0.
\end{equation}
Note that Eq. (\ref{eq27}) can be precisely obtained from similar
dispersion Eq.  (19) from \cite{Vadas} by excluding terms
comprising first and second viscosities. For simplicity, consider
IGW continuum waves with low frequencies $\sigma$ and $k_z^*\gg
1$. In this case, Equation
       \begin{equation}\label{eq28}
-{k_z^*}^2+q^2+\frac{e^{z^*}}{i}\left({k_z^*}^2+k^2\right)^2=0
\end{equation}
gives four simple roots, one of which corresponding to upward
propagating IGW wave at $z^*\rightarrow -\infty$  will be
interesting to us:
       \begin{equation}\label{eq29}
k_z^*=-\sqrt{\frac{2K^{2}}{\left( \sqrt{4ie^{z^*}K^{2}+1}+1\right)
}-k^{2}},
\end{equation}
where $K=\sqrt{q^2+k^2}$ is the full vector of a dissipationless
wave. The dissipative wave attenuation at $z^*$ may be presented
in the form:
       \begin{equation}\label{eq30}
\eta_{WKB}=\sqrt{\frac{q}{\left|k_z^*(z^*)\right|}}e^{-{\rm
Im}\left[\int^{z^*}_{\infty}k_z^*(z^{*'})dz^{*'}\right]}=\sqrt{\frac{q}{\left|k_z^*(z^*)\right|}}e^{-\Gamma}.
\end{equation}
Expression of Eq. (\ref{eq30}) may be used for estimating wave
attenuation in the real atmosphere. In the isothermal atmosphere
approximation, the integral in the exponent may be presented in
the analytical form:
      \begin{equation}\label{eq31}
\Gamma={\rm Re}\left[i\sqrt{2}K\left(2in{\rm
Ln}\frac{a+in}{a-in}+b{\rm
Ln}\frac{b-a}{b+a}-2a-i\frac{\pi}{2\sqrt{2}}\right)\right],
\end{equation}
where $n=\frac{k}{\sqrt{2}K}$;
$a=\sqrt{\frac{1}{1+\sqrt{1+4iKe^{z^*}}}-n^2}$;
$b=\sqrt{\frac{1}{2}-n^2}$. In addition to the value determined by
Formula of Eq. (\ref{eq30}), the attenuation value versus the
disturbance amplitude at the given lower height $z_n^*$ may also
be useful:
      \begin{equation}\label{eq30_a}
\eta_{WKB}(z^*,z_n^*)=\eta_{WKB}(z^*)/\eta_{WKB}(z_n^*).
\end{equation}

\begin{figure}
    \includegraphics[width=1\linewidth]{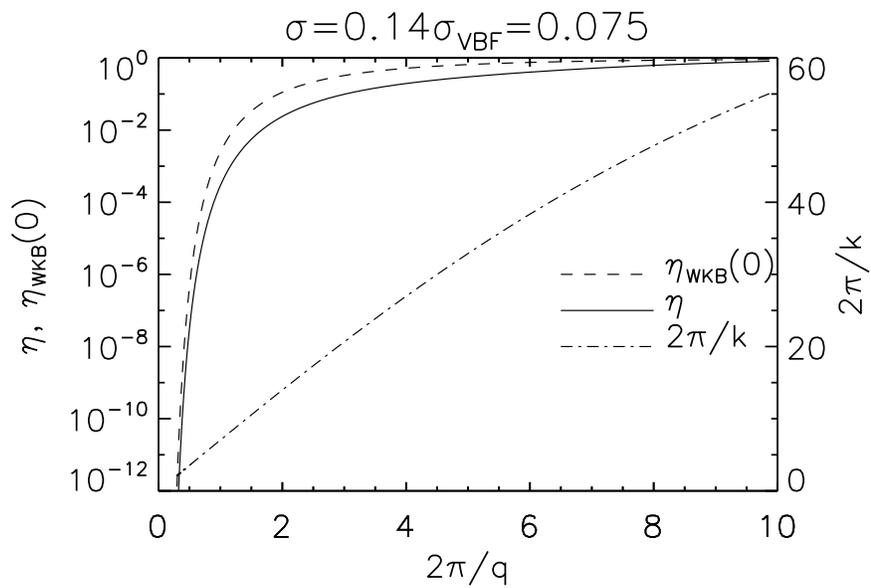}\\
  \caption{The solid line is the dependence of $\eta$ on vertical length
of the incident wave; the dashed line is the same for
$\eta_{WKB}(0)$; the dash-dotted line is the dependence of
horizontal wavelength on vertical length of the incident wave for
the selected wave solutions   (the left axis). The frequency
$\sigma$ is constant (0.14 of the Vaisala-Brent frequency
$\sigma_{VBF}$).}\label{fig2}
\end{figure}

\begin{figure}
    \includegraphics[width=1\linewidth]{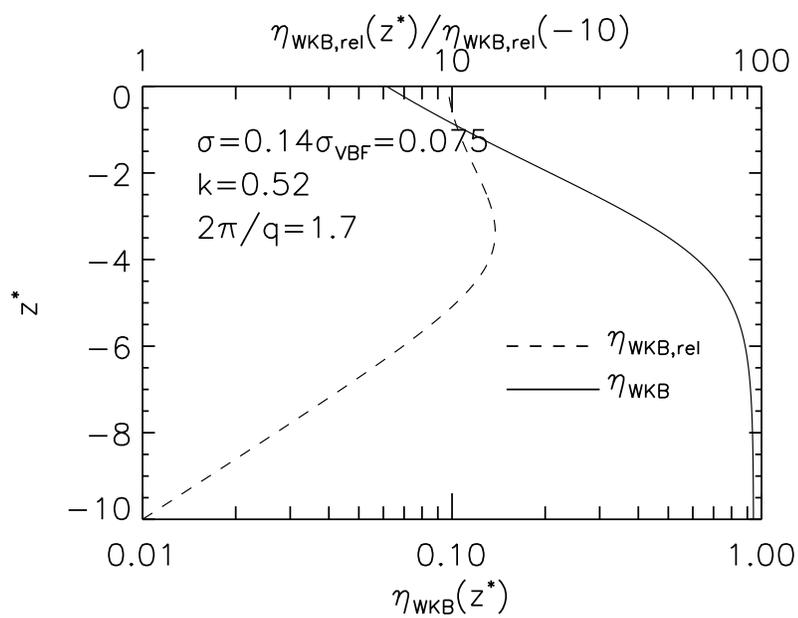}\\
  \caption{The height dependence of attenuation values  $\eta_{WKB}$ and $\eta_{WKB,rel}$ at
given wave parameters.}\label{fig3}
\end{figure}

The wave attenuation $\eta_{WKB}(0)$ expressed via Eq.
(\ref{eq31}) is comparable with similar value in Eq.(\ref{eq25})
obtained from the analytical wave solution (Fig. \ref{fig2}); as
function of $z^*$, this value qualitatively describes the
amplitude of the wave solution in the isothermal atmosphere below
$z_c$ ($z^*<0$) (Fig. \ref{fig3}).

Fig. \ref{fig2} demonstrates convergence of the wave attenuation
dependences  at decreasing vertical wave scales. This comparison
justifies validity of the introduced analytical measure of
attenuation.  The fact that the introduced new measure is
determined without limitations for any wave parameters warrants
its use as a universal wave characteristic. Alternatively, the
obtained convergence in inverse limit (large scales) may justify
applying the WKB approximation to amplitude estimates of real
long-wave disturbances, though formally this approximation is not
valid.

Fig. \ref{fig3} illustrates the quite predictable typical behavior
of the vertical distribution of wave amplitudes below $z_c$.  It
is obvious that the dissipation effect begins several scales of
the height of the atmosphere to the critical height. The
additional characteristic
$\eta_{WKB,rel}=e^{\frac{1}{2}z^*}\eta_{WKB}$ demonstrates the
required dissipation-provoked suppression of the exponential
growth of the amplitude of relative oscillations. The suppression
of the exponential growth provides, in turn, a possibility of
satisfying the linearity of disturbance at all heights.

\begin{figure}
    \includegraphics[width=1\linewidth]{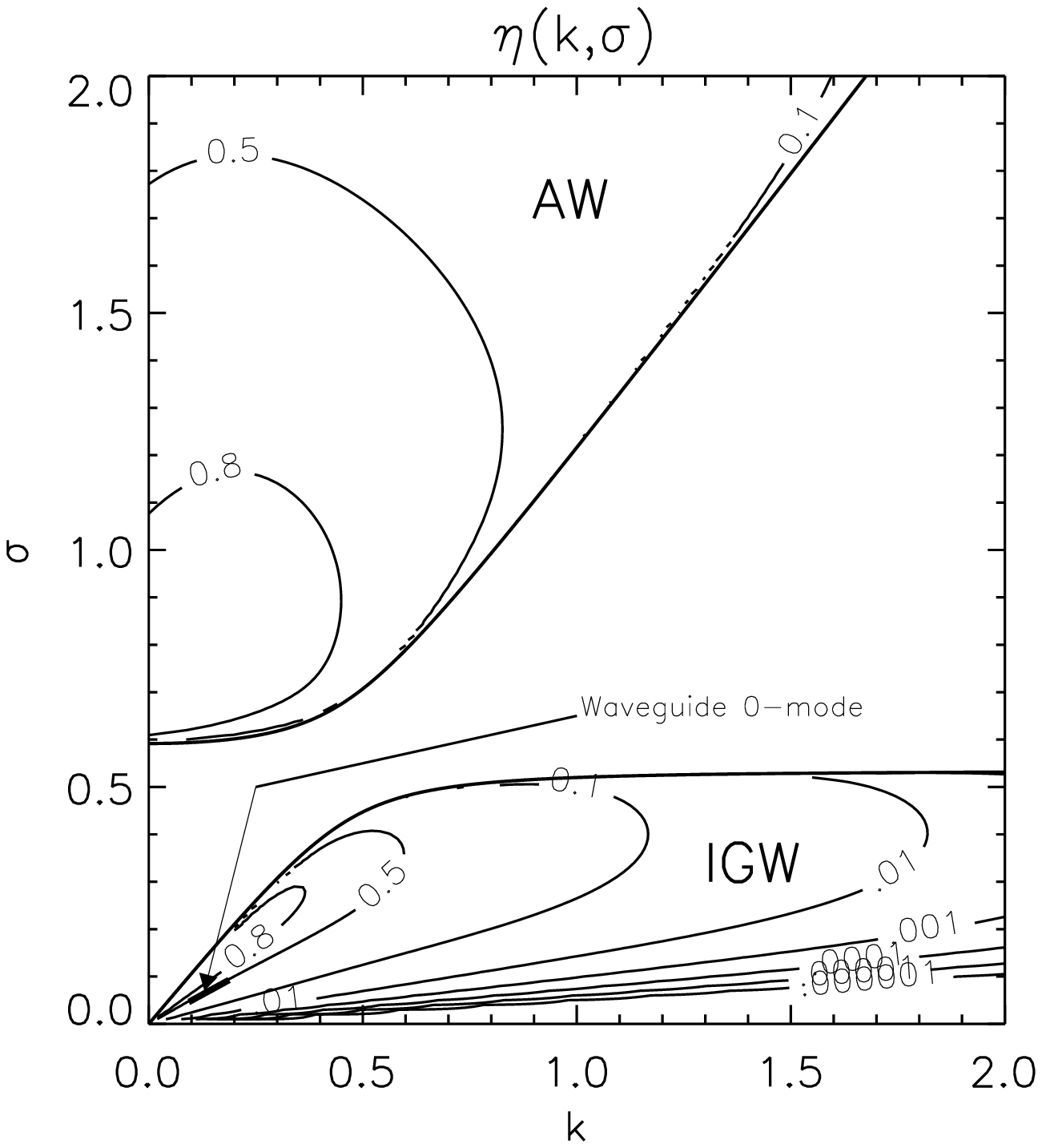}\\
  \caption{The levels of $\eta$ in the plane of $k$ and $\sigma$.}\label{fig4}
\end{figure}

\begin{figure}
    \includegraphics[width=1\linewidth]{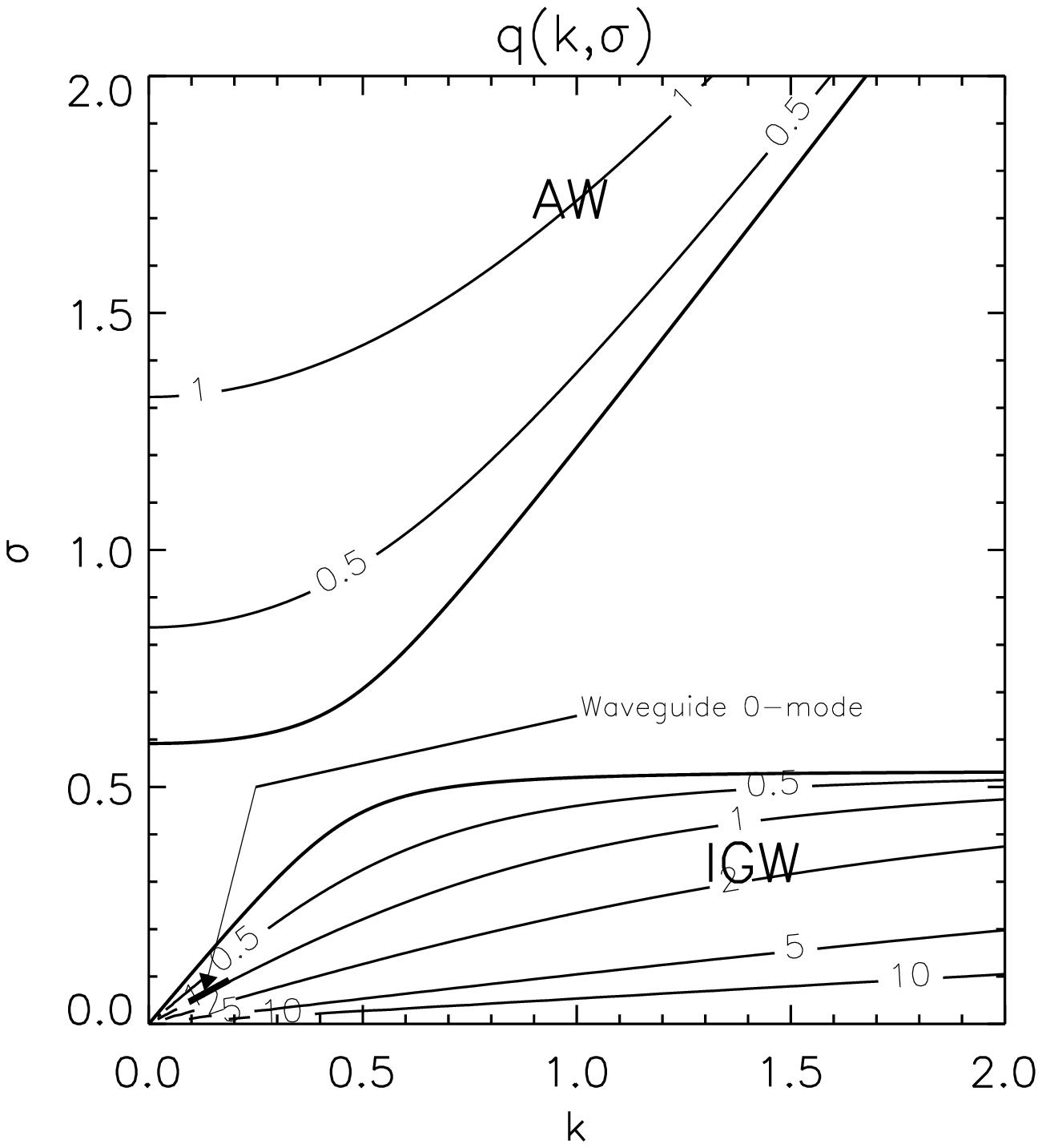}\\
  \caption{The levels of $q$ in the plane of $k$ and $\sigma$.}\label{fig5}
\end{figure}

A general picture of distribution of universal characteristic of
attenuation for internal gravity and acoustic waves is given in
Fig. \ref{fig4} which shows levels of constant values $\eta$ in
the plane of wave parameters ($k,\sigma$). For convenience of
comparison, Fig. \ref{fig5} presents levels of constant values of
the vertical wave number $q$ in the same plane. Fig. \ref{fig4}
and Fig. \ref{fig5} classify waves of weak, moderate, and strong
attenuation according to $k$ and $\sigma$ and allow us to estimate
the possibility of their penetration to heights above $z_c$.
\section{Long-distance disturbance propagation in the upper atmosphere}\label{section5}
\begin{figure}
    \includegraphics[width=1\linewidth]{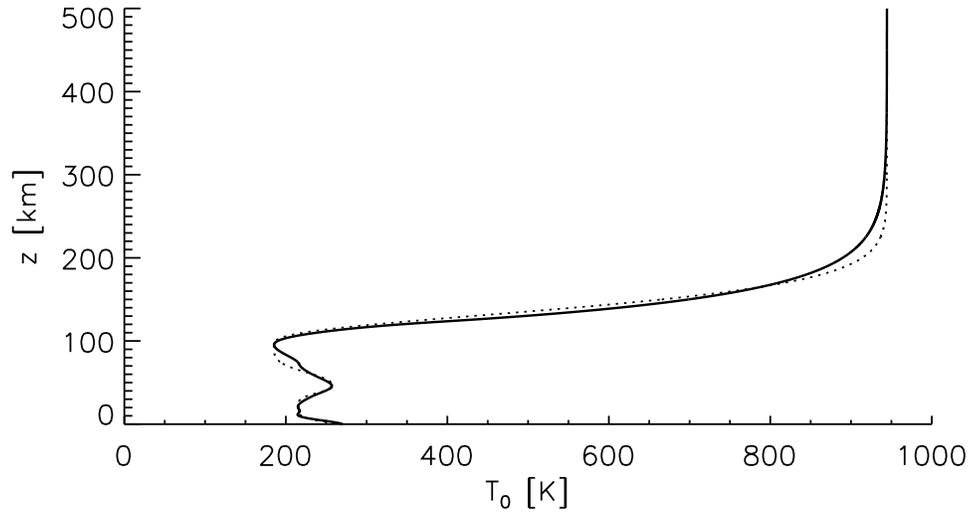}\\
  \caption{The approximate temperature dependence on height (dotted
line); the basic dependence (solid line).}\label{fig6}
\end{figure}

This section addresses a special type of long-period disturbances
which occur at ionospheric heights far from their sources. Seeing
that such waves can not be captured on their own in the upper
atmosphere (approximately isothermal), the only way to explain the
observation is to consider the waves as a result of propagation
from a waveguide located at lower heights.    A rigorous
description of waveguide propagation in the real atmosphere
(without accounting for dissipation) may be obtained, in
principle, from the solution of a boundary problem for a wave
equation which more completely accounts for the stratification of
the real atmosphere (\cite{Ost}; \cite{Pon}). For simplicity, we
will restrict ourselves to wave Eq. (3) from \cite{Pon} in its
short form for the windless atmosphere.
  \begin{equation}\label{eq32}
\begin{array}{l}
  \Psi''+U\Psi=0,  \\
  U(z)=\\
  \frac{1}{2}\left( \frac{g\rho_0}{\gamma p_0}-Q\right) ^{\prime }-\frac{1}{4}\left( \frac{g\rho_0}{\gamma p_0}-Q\right) ^{2}
  +\left(\frac{\omega^2\rho_0}{\gamma p_0}-k_x^2\right)\left[1-\left( \ln \frac{p_{0}^{1/\gamma }}{\rho _{0}}\right) ^{\prime }\frac{g}{\omega ^{2}}\right], \\
  \\
  Q=\left\{ \ln \frac{p_{0}^{1/\gamma }}{\rho _{0}\left[ \omega ^{2}-\left( \ln \frac{p_{0}^{1/\gamma }}{\rho_0}\right) ^{\prime }g\right] }\right\} ^{\prime }.
\end{array}
\end{equation}
Here, wave function $\Psi$ is related to the pressure disturbance
through \\
 $p=\Psi\left\{\rho_0p_0^{-1/\gamma}\left[\omega^2-\left(\ln\frac{p_0^{1/\gamma}}{\rho_0} \right)^{\prime }g \right] \right\}^{1/2}
 \exp\left(-\frac{1}{2}\int^z_0\frac{g\rho_0}{\gamma p_0}d\xi \right)e^{-i\omega t+ik_xx}$. \\
The possibility of waveguide propagation depends on the presence
of limited height intervals with positive values of $U$-function
which may be interpreted as squared vertical wave vector $k_z$ in
the WKB approximation. Here, we will compare two methods for
describing wave solutions with penetration, using the rigorous
solution of the boundary wave problem for Eq. (\ref{eq32}) and the
WKB approximation. The WKB description for the wave phenomena with
vertical scales comparable with atmospheric irregularities is,
strictly speaking, rather conditional. Nevertheless, we will
demonstrate that this description is quite sufficient to obtain
dispersive properties of waveguide solutions and is more
convenient for their physical interpretation. From Eq.
(\ref{eq32}) follows that constructing the $U$-function requires
continuity of not only a temperature function but also of its
first and second derivatives. The initial temperature distribution
(Fig. \ref{fig1}) has a certain amount of points with derivative
discontinuities that leads to unphysical $U$-function  jumps.  For
the sequel, we will therefore approximate the temperature
dependence to a set of reference functions satisfying necessary
conditions of smoothness:
   \begin{equation}\label{eq33}
\begin{array}{l}
T(z>430 )=944.4, \\
T(95.3<z\leq 430 )=
\\
\left(\left[\cos\left(\frac{\pi}{2}\left(\frac{430-z}{430-95.3}\right)^6\right)\right]^3-1\right)(944.4-185.4)-944.4,\\
\\
T(46<z\leq 95.3 )=\\
\left(\left[\cos\left(\frac{\pi}{2}\left(\frac{95.3-z}{95.3-46}\right)^2\right)\right]^3-1\right)(257-185.4)-257,\\
\\
T(20<z\leq 46 )=\\
\left(\left[\cos\left(\frac{\pi}{2}\left(\frac{z-20}{46-20}\right)^2\right)\right]^3-1\right)(215.1-257)-215.1,\\
\\
T(0<z\leq 20 )=\\
2\left(\left[\cos\left(\left(\frac{20-z}{20}\right)^2\arccos\left(0.5^{3/2}\right)\right)\right]^3-1\right)(215.1-270.1)-215.1.
\end{array}
\end{equation}
The comparison of the new temperature dependence on height with
the basic dependence is shown in Fig. \ref{fig6}.

Discuss the U-function profile (Fig. \ref{fig6}) for arbitrarily
selected wave parameters $\omega$ and $k$ ($T_w=2\pi/\omega=90
 \ {\rm min}; \lambda_{hor}=2\pi/k_x=1390  \ {\rm km}$).
It is evident that the waveguide may be located below $z_1$. If
the waveguide solution is implemented by the $U$-profile, the
upper locking wall of the waveguide is a region of negative
$U$-values: $z_1<z<z_2$ . The same region, in turn, is also a
barrier through which a part of wave energy may escape. Above
$z_2$ is again a region of wave propagation extending up to the
heights we are interested in. A distinguishing characteristic of
the problem in hand is a strong variation in form and $U$-values,
depending on wave parameters. Values of $z_1$ and $z_2$ change
too.
\subsection{Boundary wave problem (BWP) for wave modes}\label{section5.1}
It is most convenient to solve the boundary problem for Eq.
(\ref{eq32}) via the corresponding nonlinear Riccati equation:
       \begin{equation}\label{eq1kr}
G'-G^2U-1=0,
\end{equation}
Where $G$ is related to $\Psi$ through
        \begin{equation}\label{eq2kr}
G\Psi'=\Psi.
\end{equation}
The $G$-function of the waveguide solution must meet top and
bottom boundary conditions. At the top
($z=z_\infty\rightarrow+\infty$), the $G$-function must fit an
upward IGW:
         \begin{equation}\label{eq3kr}
G(z_\infty)=1, \ G'(z_\infty)=i/\sqrt{u(z_\infty)}.
\end{equation}
In the numerical implementation, $z_\infty$ was taken to be 270 km
above which, in our model, the $U$-function is constant. At the
bottom ($z=0$), we impose a requirement:
         \begin{equation}\label{eq4kr}
G(0)=0.
\end{equation}
This requirement, according to Eq. (\ref{eq2kr}), is equivalent to
the condition of equality to zero of the $\Psi$-function.

The numerical solution of Cauchy problem Eqs. (\ref{eq1kr}),
(\ref{eq3kr}) in the $G$-function allows us to derive a complex
dispersion equation
         \begin{equation}\label{eq5kr}
D(\omega,k_x)=G(0,\omega,k_x).
\end{equation}
Formally, we can solve Eq. (\ref{eq5kr}) by assuming the first or
the second argument of dispersion $D$-function to be real. In the
former case, we will have modes attenuating (owing to the
nonhermiticity of the problem) in the horizontal direction of
propagation; in the latter case, modes attenuating in time. In
this paper, we analyze modes only with real values of frequency
$\omega$. A vertical spatial structure of the mode (for the pair
of dispersive values $\omega$ and $k_x$ satisfying Eq.
(\ref{eq5kr})) can be obtained by solving numerically the Cauchy
problem for Eq. (\ref{eq32}) with the initial condition $\Psi=0$,
$\Psi'=1$ corresponding to boundary condition of Eq.(\ref{eq4kr})
for the $G$-function.
\subsection{WKB analysis of waveguide solution}\label{section5.2}
In the general case, the condition of waveguide locking with
energy leakage in the WKB approximation may be represented by a
modified Bore-Sommerfeld condition of quantization (MBSCQ) with
complex turning points:
         \begin{equation}\label{eq34}
         \begin{array}{l}
\int_C\sqrt{U(z)}dz\approx\pi\left(\frac{1}{2}+n\right)+i\exp\left[-2\int_{z_1}^{z_2}\sqrt{|U_0(z)|dz}\right]\equiv\\
\pi\left(\frac{1}{2}+n\right)+iE, \ n=0,1,... \ .
\end{array}
\end{equation}
Condition of Eq. (\ref{eq34}) gives  dispersion ratios between
real frequencies $\omega$ and complex wave numbers $k_x$ with  a
small imaginary part accounting for the degree of horizontal
attenuation of the waveguide mode $n$. The integration contour $C$
of the integral on the left-hand side of Eq. (\ref{eq34}) begins
from $z_0$ and ends at a complex turning point $z_{c1}$ close to
the real turning point $z_1$($U(z_1,\omega,{\rm Re}k_x)$).
Besides, with inner (complex) turning points, we assume that the
$C$-contour also passes through these points (in our calculations,
for simplicity, sections of the $C$-contour with ${\rm Re}U(z)<0$
are ignored). The integral in the exponent argument on the
right-hand side of Eq. (\ref{eq34}) is assumed to be real; the
$0$-index of the $U$-function in the integrand means that it is a
function of the real part $k_x$ and real $z$:
$U_0(z)=U(z,\omega,{\rm Re}k_x)$. Solve Eq. (\ref{eq34}), using
the perturbation theory:
          \begin{equation}\label{eq35}
\int_{0}^{z_1}\sqrt{|U_0(z)|dz}-\pi\left(\frac{1}{2}+n\right)=0
\end{equation}
for the selected real $\omega$, find a real root of $k_{0x}$ in
Eq. (\ref{eq35}); next, insymbols
          \begin{equation}\label{eq36}
k_x=k_{0x}(1+i\delta), \ U(z)=u_0(z)+k_{0x}i\delta\frac{\partial
}{\partial k_x}U_0;
\end{equation}
By substituting Eq. (\ref{eq36}) to Eq. (\ref{eq34}) and
accounting for the complexity of turning points $z_{cj}$ of the
integration contour of the integral on the left-hand side of Eq.
(\ref{eq34}), we obtain an equation for complex addition of the
horizontal wave vector:
          \begin{equation}\label{eq37}
\delta\frac{1}{2}\int_0^{z_1}\frac{k_{0x}}{\sqrt{U_0}}\left(\frac{\partial}{\partial
k_x}U_0\right)dz+\delta^{3/2}\frac{2}{3}e^{3i\pi/4}k_{0x}^{3/2}\sum_j\left[\frac{(\partial
 U_0/\partial k_x)^{3/2}}{|\partial U_0/\partial z|}\right]_{z=z_j}=E.
\end{equation}
We have, thus, three algebraic roots of cubic Eq. (\ref{eq37}) for
$\delta^{1/2}$. We choose one of them that reflects physical
attenuation of the wave mode.
\subsection{Numerical calculations of characteristics of internal gravity waveguide modes}\label{section5.3}
First, we have established that there is only one nodeless
waveguide mode (with $n=0$) in the selected model atmosphere in
the frequency range that corresponds to large-scale TIDs . This
was demonstrated by both the algorithms described in Subsections
\ref{section5.1} and \ref{section5.2}. The dispersion curve of the
$0$-mode is shown by a segment of the thick curve in the plane of
wave dimensionless parameters ($\sigma,k$) in Figs. \ref{fig4} and
\ref{fig5}. For a more detailed analysis of propagation
characteristics of the waveguide mode,  Fig. \ref{fig8} gives: \\
-- the dispersion dependence of horizontal wavelength on
oscillation period (the dotted curve is the dependence obtained
from BWP; the solid curve is the dependence obtained from
MBSCQ);\\
-- asymptotic characteristics of the wave leakage to the upper
atmosphere: full phase velocity (dashed curve); vertical group
velocity (dash-dotted curve); vertical wavelength (dash-dot-dotted
curve). \\
Fig. \ref{fig9} presents estimated characteristics of the
horizontal attenuation: the dotted curve is the characteristics
obtained from BWP; the solid curve, from MBSCQ.

Note the most important points: \\
--   The case of model atmosphere, we analyzed, showed that there
is only one mode. Seeing that the chosen time of the model for the
geographic localization considered   corresponds most often to
moments of detection of ionospheric disturbances, we may assume
that the implementation of the conditions (somewhere) for two or
more modes is most likely to be extremely rare or impossible at
all. \\
--   It is important that the obtained dispersion dependence
fairly faithfully reproduces the observed ratio of horizontal
scales to periods of TIDs and gives reasonable estimates for their
full phase velocities obtainable in measurements. This result
furnishes convincing proof of their inextricable connection with
the waveguide propagation of IGW. \\
-- The two approaches (BWP and MBSCQ) show a perfect match for
each other in phase characteristics and a quite satisfactory match
in attenuation sufficient for very long-distance propagation of
waveguide disturbances.  The analysis of the structure of the
rigorous waveguide solution (Fig. \ref{fig10}) demonstrates that,
despite the complex behavior of the $U$-function, the height
dependence of the solution is smooth with a typical scale of
change far exceeding that of  $U$ change. It is even more
surprising that the quasiclassical description, despite its formal
invalidity, provides results which are very close to the exact
solution. Thus, we verify practicality for the quasiclassical
description of the problem in hand.\\
It is interesting to note a property which is likely to be
specific only for IGW  waveguides. The dependence in Fig.
\ref{fig9} shows a high Q factor of oscillations relative to the
characteristic of horizontal attenuation in spite of the fact
that, in the opacity barrier $[z_1,z_2]$, the amplitude of the
rigorous solution (Fig. \ref{fig10}) decreases slightly. The
parameter $\sqrt{E}$, determined in Eq. (\ref{eq34}), is of order
of 0.41. For an acoustic waveguide, its horizontal attenuation
characteristic would be estimated as $0.41^2$.  For IGW, the
multiplier of the first term in Eq. (\ref{eq37}) gives a value of
order of 30 (for AGW, it would be $\sim 1$) on the whole
dispersion curve. This factor causes a very weak attenuation of
mode. On the other hand, the physical substantiation of this may
be the low vertical group velocity of the upward propagating wave
penetrating through the opaque region (Fig. \ref{fig9}). The slow
energy leakage from a waveguide causes the slow waveguide
attenuation.
\subsection{Analyzing the height dependence of amplitude-phase characteristics of the wave penetrating to the upper "dissipative" atmosphere}\label{section5.4}
For a qualitative relation of the classical upward propagating
wave with its subsequent modification under the action of
dissipation, we will consider two propagation regions: $z_2<z<z_c$
and $z>z_c$. If the disturbance amplitude is set $1$ at $z=z_1$,
we can estimate the relation between the amplitude of the
waveguide solution and the amplitude of the upward propagating
wave at $z_c$: If we ignore the multiplier of the exponential
growth --
          \begin{equation}\label{eq38}
\mu(z_c;\lambda_{gor},T_w)=\sqrt{E(\lambda_{gor},T_w)}\eta_{WKB}(z_2,z_c;\lambda_{gor},T_w);
\end{equation}
In view of the exponential growth  --
  \begin{equation}\label{eq39}
\mu_{rel}(z_c;\lambda_{gor},T_w)=\mu(z_c;\lambda_{gor},T_w)\cdot\left(\frac{\rho_0(z_2)}{\rho_0(z_c)}\right)^{1/2}.
\end{equation}
Here, $\lambda_{gor}$ and $T_w$ are respective horizontal
wavelength and wave period on the dispersion curve of the
waveguide $0$-mode; the first multiplier on the right-hand side of
Eq. (\ref{eq38}) corresponds to the quasiclassical estimate of the
decrease in amplitude in the opacity barrier according to
definition Eq. (\ref{eq34}); the second multiplier on the
right-hand side of Eq. (\ref{eq38}) gives a change of the
amplitude under the action of dissipation according to dissipative
WKB estimate Eqs. (\ref{eq30}), (\ref{eq30_a}). Corresponding
plots of these values are given in Fig. \ref{fig11}.

As was shown in Section \ref{section3} , at $z>z_c$, we can find
the exact dissipative wave solution from  (\ref{eq23}),
(\ref{eq12}). This is demonstrated by Fig. \ref{fig12} and Fig.
\ref{fig13} for one pair of wave parameters $\lambda_{gor}$,$T_w$
on the dispersion curve of $0$-mode of a wave with $T_w=1.5 \ {\rm
h}$. Amplitude characteristics of all wave physical quantities in
Fig. \ref{fig12} are normalized to the amplitudes in accordance
with equality 1 of the amplitude of the relative disturbance of
density at $z=z_c$. For convenience, vertical lines show initial
amplitudes of the corresponding classical dissipationless wave at
$z=z_c$.  Real changes of classical dissipationless amplitudes of
IGW are not presented here because of their unlimited exponential
growth (several orders of their values in the given height
interval). We can see that the relations between dissipative
amplitude values of the exact solution almost exactly coincide
with the similar relations of the classical solution (unchanged
with height) at $z=z_c$. The relations between amplitudes of the
dissipative solution change with height significantly and
demonstrate the predicted stop of the exponential growth with
height and their subsequent gradual decrease (the beginning of
which can be observed for all amplitudes in the upper height
range). A similar behavior can be observed in relations between
phases of all values in Fig. \ref{fig13}, i.e. they change with
height significantly, coinciding with relations between phases of
the classical solution at $z=z_c$*.  The fact that the relations
between amplitudes and between phases correspond to the classical
solution counts in favor of the dissipative WKB description up to
$z_c$.
\section{Conclusion}\label{section6}
We have demonstrated that the physical interpretation of
disturbances in the upper atmosphere (especially at ionospheric
heights) requires a rigorous wave description accounting for the
dominant dissipation effect. Beginning from a critical height
($z_c$) depending on an oscillation period, behavior of wave
disturbances cannot be described in any form of the WKB
approximation.   Below this height, this approximation can be
successfully exploited in the part of the upper atmosphere with
dissipative corrections and then, in its classical
(dissipationless) form, can satisfactorily describe wave
propagation in the real, significantly stratified atmosphere. In
this case, for low-frequency oscillations of the IGW range, the
WKB description remains valid (despite its formal incorrectness)
for waves vertical scales of which are comparable to or larger
than scales of vertical irregularity of a medium. This statement
is based on the remarkable coincidence we obtained between results
of calculations of the zero waveguide mode. Waves with vertical
scales comparable with the height of the atmosphere were shown to
be able to reach a critical height and propagate further to higher
ionospheric heights. Under the action of dissipation, these waves
are considerably transformed: an intense exponential growth stops
followed by a slow amplitude decrease allowing the wave to retain
its linear features. This conclusion is based on our analysis of
the rigorous hypergeometric solution of the wave problem for the
isothermal thermal-conductivity atmosphere. We have examined a
phenomenological hybrid model for physical interpretation of
formation and long-distance propagation of TIDs. This model
assumes that manifestations of ionospheric disturbances are caused
by the interaction between a charged component and a wave
disturbance in the neutral atmosphere, which penetrates from a
lower waveguide. The construction of appropriate waveguide
solutions showed that there may be only one nodeless $0$-mode. We
studied its dispersion features and estimated horizontal
attenuation demonstrating the possibility for long-distance
horizontal propagation of the mode. The obtained dispersion
dependence of horizontal wave scale on wave period agrees well
with known spatio-temporal characteristics of TIDs. The estimated
phase velocities of the upward propagating wave also closely match
the observed phase PIV velocities(\cite{Medv}; \cite{Rat}).

\textbf{Acknowledgment}

This work is supported by the RFBR (11-05-00698-a).

\bibliographystyle{elsarticle-harv}



\newpage
\begin{figure}
    \includegraphics[width=1\linewidth]{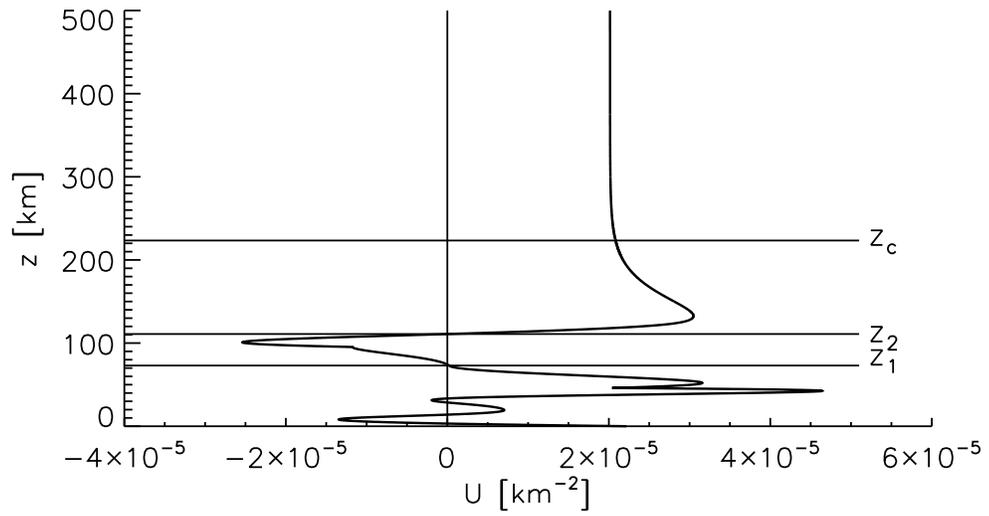}\\
  \caption{The characteristic height distribution of $U$-function.}\label{fig7}
\end{figure}
\begin{figure}
    \includegraphics[width=1\linewidth]{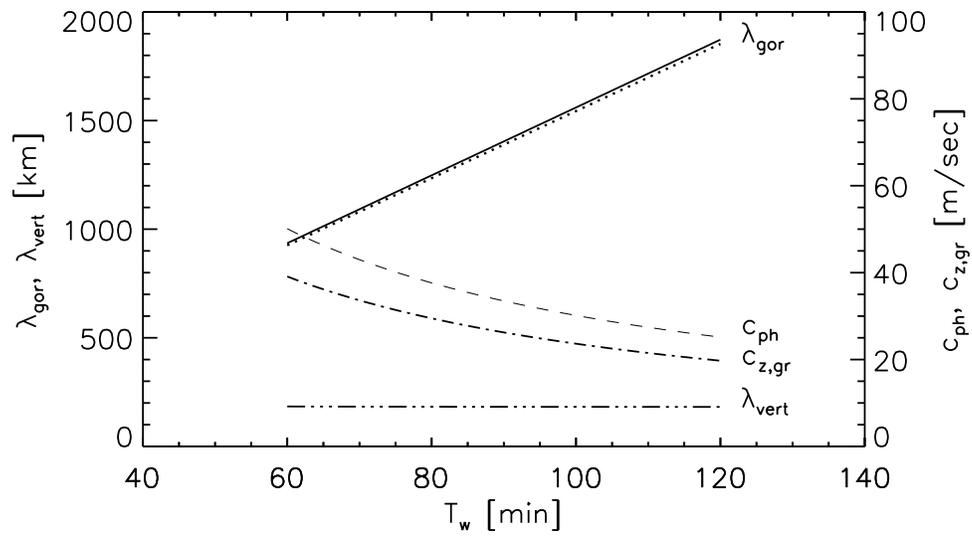}\\
  \caption{The waveguide characteristics of $0$-mode:  horizontal
wavelength from MBSCQ (solid line); horizontal wavelength from BWP
(dotted line); full phase velocity of the upward propagating wave
(dashed line); vertical group velocity of the upward propagating
wave (dash-dotted line); vertical length of the  upward
propagating wave (dash-dot-dotted line).}\label{fig8}
\end{figure}
\begin{figure}
    \includegraphics[width=1\linewidth]{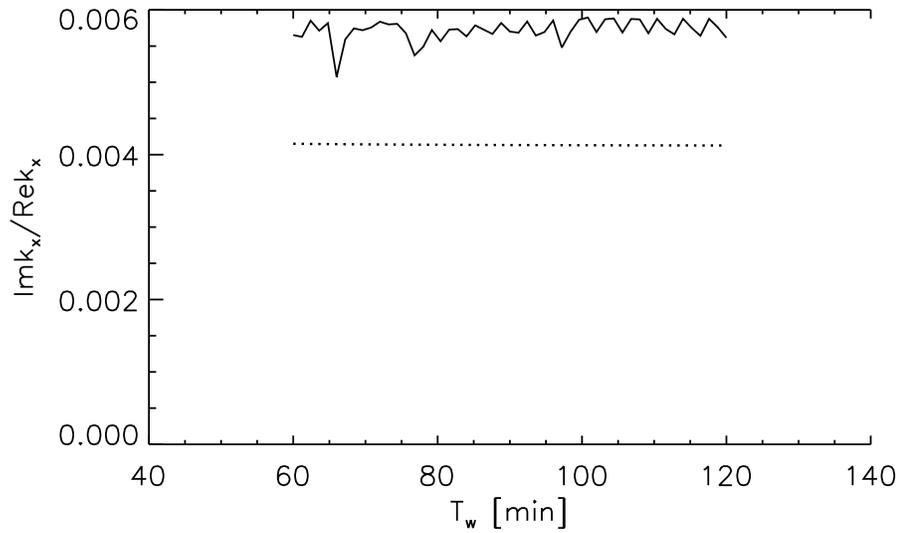}\\
  \caption{The waveguide characteristics of $0$-mode: horizontal
attenuation characteristic from MBSCQ (solid line); horizontal
attenuation characteristic from BWP.}\label{fig9}
\end{figure}
\begin{figure}
    \includegraphics[width=1\linewidth]{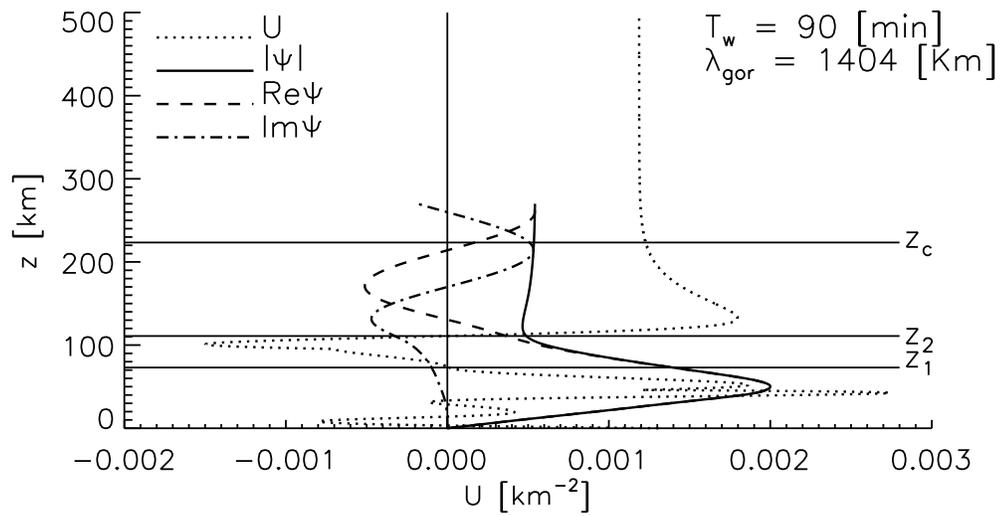}\\
  \caption{An example of the vertical structure of the waveguide
solution for $0$-mode.}\label{fig10}
\end{figure}
\begin{figure}
    \includegraphics[width=1\linewidth]{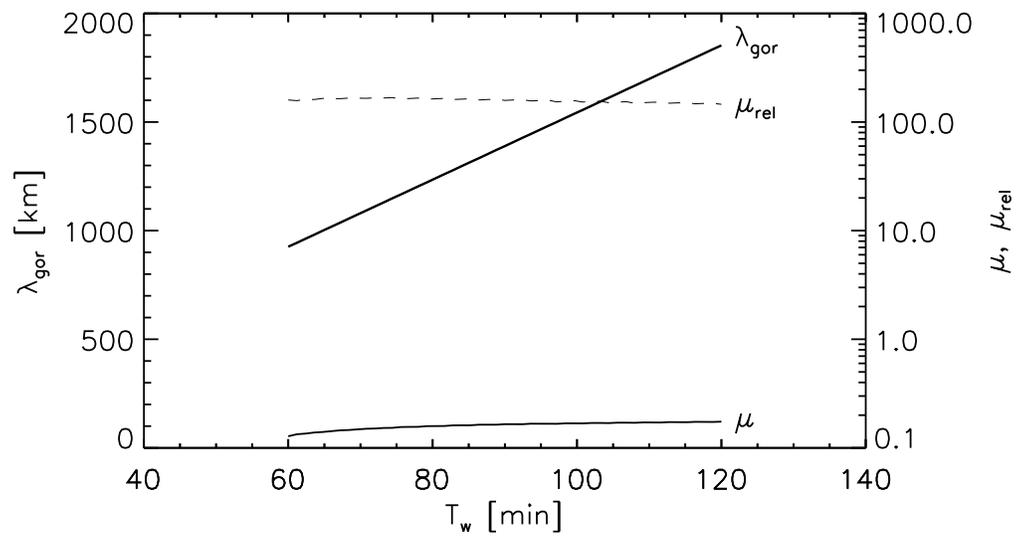}\\
  \caption{The characteristics of the ratio of the upward
propagating wave at the critical height to the amplitude of the
waveguide solution at the low height of the waveguide
opacity.}\label{fig11}
\end{figure}
\begin{figure}
    \includegraphics[width=1\linewidth]{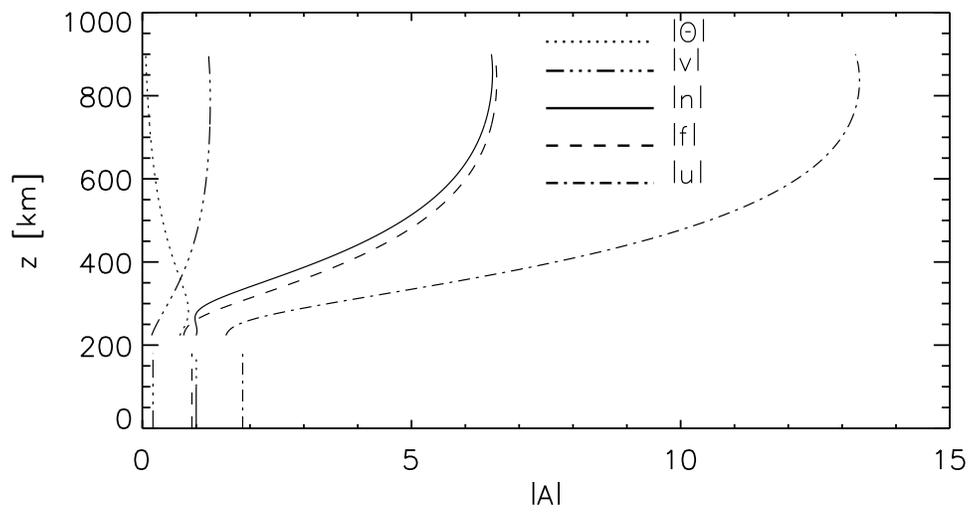}\\
  \caption{The amplitude characteristics of wave oscillation above
the critical height at dominant dissipation.}\label{fig12}
\end{figure}
\begin{figure}
    \includegraphics[width=1\linewidth]{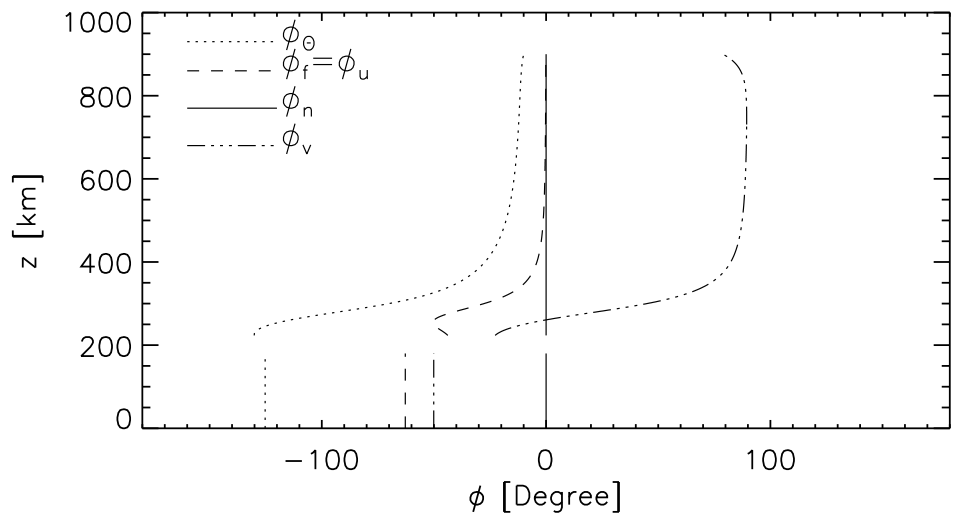}\\
  \caption{The phase
characteristics of wave oscillation above the critical height at
dominant dissipation.}\label{fig13}
\end{figure}

\end{document}